\documentclass[final,british,envcountsect,runningheads,a4paper]{llncs}

\makeatletter
\newcommand*{\installoption}[2][llncs]{
  \expandafter\newif\csname if#2\endcsname
  \@ifclasswith{#1}{#2}{
    \csname#2true\endcsname
  }{}
}
\newcommand*{\obeyoption}[2]{
  \csname if#2\endcsname\else\csname #1false\endcsname\fi
}
\newcommand*{\excludeoption}[2]{
  \csname if#2\endcsname\csname #1false\endcsname\fi
}
\makeatother

\installoption{final}
\installoption{review}
\excludeoption{review}{final}
\installoption{draft}
\installoption{showframe}
\obeyoption{showframe}{draft}
\installoption{appendix}
\obeyoption{appendix}{review}
\installoption{spacecrimes}
\obeyoption{spacecrimes}{review}

\usepackage{xparse}
\usepackage{etoolbox}
\usepackage{environ}
\usepackage[final]{graphicx}
\usepackage{realboxes}
\usepackage{circledsteps}

\makeatletter
\newcommand{\IfDefTF}[3]{\@ifundefined{#1}{#3}{#2}}
\newcommand{\IfDefT}[2]{\IfDefTF{#1}{#2}{}}
\newcommand{\IfDefF}[2]{\IfDefTF{#1}{}{#2}}
\newcommand{\IfLabelExistsTF}[3]{\IfDefTF{r@#1}{#3}{#2}}
\newcommand{\IfLabelExistsT}[2]{\IfLabelExistsTF{#1}{#2}{}}
\newcommand{\IfLabelExistsF}[2]{\IfLabelExistsTF{#1}{}{#2}}
\makeatother

\usepackage[T1]{fontenc}
\usepackage[utf8]{inputenc}
\usepackage{microtype}

\ifshowframe
\AtEndPreamble{%
  \usepackage{showframe}%

}
\fi

\usepackage{babel}

\usepackage{xcolor}

\definecolor{datacolor}{rgb}{1,.949,.8}
\definecolor{apicolor}{rgb}{1,.9,.8}
\definecolor{behaviourcolor}{rgb}{.854,.91,.995}
\definecolor{aspectscolour}{rgb}{.835,.91,.831}

\usepackage{float}
\usepackage{subcaption}
\usepackage[section]{placeins}

\usepackage[inline]{enumitem}

\usepackage{array}
\usepackage{tabularx}
\usepackage{xtab}
\usepackage{multirow}
\usepackage{booktabs}
\usepackage{colortbl} 

\usepackage[hidelinks,final]{hyperref}

\usepackage[capitalise,nameinlink,noabbrev]{cleveref}

\usepackage{url}
\usepackage{xurl} 
\newcommand{\furl}[1]{\footnote{\url{#1}}} 

\usepackage{fmtcount}
\usepackage{numprint}
\npthousandsep{,}
\npdecimalsign{.}
\newcommand{\printNo}[1]{%
	\ifnum #1 > -1
		\ifnum #1 < 11
			\numberstringnum{#1}%
		\else
			\numprint{#1}%
		\fi
	\else
		\numprint{#1}%
	\fi%
}

\usepackage{listings}
\definecolor{lstbackground}{RGB}{255,255,255}
\definecolor{lstcomment}{RGB}{0,125,0}
\definecolor{lstnewkeyword}{RGB}{0,125,0}
\definecolor{lststring}{RGB}{42,0,255}

\definecolor{lstkeyword}{RGB}{0,0,0}
\definecolor{lstspecialfeature}{HTML}{BF4040}
\definecolor{lstdatatyperule}{HTML}{0000d6}
\definecolor{lstannotation}{HTML}{6e2c00}
\definecolor{lstspecialkeyword}{RGB}{171,48,7}
\definecolor{lstfeature}{HTML}{BF4040}
\definecolor{lstapicomment}{HTML}{CE5C00}
\definecolor{lstapiparameter}{HTML}{7F7F9F}
\definecolor{lsttasktag}{HTML}{7F9FBF}
\definecolor{lstatlcontextkeyword}{RGB}{154, 88, 176}
\lstdefinelanguage{lemma}{
	morekeywords={list, string, structure, import, datatypes, from, as, functional, microservice, microservices, @endpoints, protocols, sync, data, formats, default, with, format, async, public, interface, out, deployment, technologies, operation, environments, container, technology, environment, deploys, values, basic, endpoints, @sync, @async, int, double, required, long, version, context, float, date, in, function, procedure, immutable, hide, types, boolean, primitive, type, byte, short, char, based, on, service, aspects, for, fields, parameters, selector, aspect, true, false, internal, extends, enum, compatibility, matrix, unspecified, infrastructure, operations, properties, used, by, nodes, is, fault, complex, interfaces, utility, op, return, domainOperations}
}
\lstdefinestyle{lemma}{	
	language=lemma,
	morecomment=[s]{/*}{*/},
	morecomment=[l]{//},
	moredelim={[is][\textcolor{black}]{$$}{$$}},
	moredelim={[is][\textcolor{darkgray}]{\%\%}{\%\%}},
	moredelim={[is][\bfseries\textcolor{lstcomment}]{§§}{§§}},
	moredelim={[is][\bfseries\textcolor{lstnewkeyword}]{\#}{\#}},
	moredelim={[is][\textcolor{lstapicomment}]{\#\#}{\#\#}},
	moredelim={[is][\itshape\textcolor{lstapiparameter}]{||}{||}},
	moredelim={[is][\bfseries\textcolor{lstkeyword}]{§}{§}},
	morestring=[b]",
	morestring=[b]'
}
\lstdefinelanguage{jolie}{
	morekeywords={type, string, length}
}
\lstdefinestyle{jolie}{	
	language=jolie,
	morecomment=[s]{/*}{*/},
	moredelim={[is][\textcolor{black}]{$$}{$$}},
	moredelim={[is][\textcolor{darkgray}]{\%\%}{\%\%}},
	moredelim={[is][\bfseries\textcolor{lstcomment}]{§§}{§§}},
	moredelim={[is][\bfseries\textcolor{lstnewkeyword}]{\#}{\#}},
	moredelim={[is][\textcolor{lstapicomment}]{\#\#}{\#\#}},
	moredelim={[is][\itshape\textcolor{lstapiparameter}]{||}{||}},
	moredelim={[is][\bfseries\textcolor{lstkeyword}]{§}{§}},
	morestring=[b]",
	morestring=[b]'
}

\usepackage[disable, colorinlistoftodos, textwidth=2.0cm]{todonotes}
\newcounter{FRToDoCounter}

\newcounter{FMToDoCounter}

\title{Jolie \& LEMMA: Model-Driven Engineering and Programming Languages Meet on Microservices\thanks{\textls[-20]{Work partially supported by Independent Research Fund Denmark, grant no.~0135-00219.}}}
\titlerunning{MDE and Programming Languages Meet on Microservices}

\author{
    Saverio Giallorenzo\inst{1} \and
    Fabrizio Montesi\inst{2} \and
    Marco Peressotti\inst{2} \and
    Florian Rademacher\inst{3} \and
    Sabine Sachweh\inst{3}
}

\institute{
	Università di Bologna, Italy and INRIA, France
	\email{saverio.giallorenzo@gmail.com}
	\and
	University of Southern Denmark
	\email{\{fmontesi,peressotti\}@imada.sdu.dk}
	\and
	University of Applied Sciences and Arts Dortmund
	\email{\{florian.rademacher,sabine.sachweh\}@fh-dortmund.de}
}

\authorrunning{S. Giallorenzo et al.}

\begin{document}

\maketitle

\begin{abstract}
In the field of microservices, Model-Driven Engineering (MDE) has emerged as a
powerful methodology for architectural design. Independently, the community of programming languages has investigated new linguistic abstractions for effective microservice development.
Here, we present the first preliminary study of how the two approaches
can cross-pollinate, taking the LEMMA framework and the Jolie programming
language as respective representatives.
We establish a common ground for comparing the two technologies in terms of metamodels, discuss practical enhancements that can be derived from the comparison, and present some directions for future work that arise from our new viewpoint.
\end{abstract}

\section{Introduction}
\label{sec:introduction}

In microservices, applications emerge as compositions of
independently-executable components (\emph{microservices}, or briefly,
services), which communicate via message passing~\cite{Dragoni2017}. Building
microservice systems poses a series of challenges for both design and
development, which has motivated two prolific strands of research.

On the side of design, Model-Driven Engineering (MDE)~\cite{F03} has become a prominent
methodology for the specification of service architectures
\cite{Ameller2015}. Frameworks such as LEMMA, MicroBuilder, and MDSL offer
modelling languages to
design service components that abstract from concrete implementations~\cite{Rademacher2020,Terzic2018,Kapferer2020b}.

On the side of development, new linguistic abstractions for programming languages are emerging as powerful tools to effectively express the
configuration and coordination of microservices.
Ballerina and Jolie are examples of such languages~\cite{Oram2019,Montesi2014}.
In particular, Jolie incorporates ideas from process calculi to ease the
programming of workflows and it offers ``polyglot'' constructs to integrate
services written in foreign languages (e.g., Java)~\cite{M16,Montesi2014}.

So far, results on microservices by the MDE and programming communities have evolved prolifically, yet separately.
This is unfortunate since previous research showed great potential in combining
programming language and MDE techniques~\cite{Erdweg2012,Butting2018,Deantoni2016}. 
In part, we deem this phenomenon due to the few opportunities the two
communities have to interact.
Case in point, the authors come from the different two communities and met only recently, at the last two editions of the International Conference on
Microservices (an event organised specifically to bridge sub-communities of traditional fields that share an interest in microservices).
Seminars from both parts evidenced that MDE methodologies and programming languages for microservices share a common conceptual foundation that has never been properly made precise nor leveraged \cite{Fernando2020,Rademacher2020d}.

This article is the first step towards bridging conceptually MDE frameworks and programming languages for microservices. As grounding, we take LEMMA~\cite{Rademacher2019,Rademacher2020} and Jolie~\cite{Montesi2014} as respective representatives of the two approaches.

The main challenge is that MDE frameworks come with specifications---like LEMMA's metamodels~\cite{Rademacher2019,Rademacher2020}---that are distant from those given for programming languages---some parts of Jolie are described by using process calculi~\cite{GLMZ09,MC11}, and for others there is a reference implementation~\cite{M10,Jolie2020}.
To address this, we develop the first conceptual metamodel of the Jolie language, drawing from our experience with its formalisations~\cite{GLMZ09,MC11,DGGMM12} and reference implementation~\cite{M10,Jolie2020}.

Having metamodels for both Jolie (from this paper) and LEMMA (from~\cite{Rademacher2019,Rademacher2020}) allows for comparing them.
We identify some key shared concepts and differences. Interestingly, the differences are complementary perspectives on common concerns, providing fertile ground for future evolutions of both approaches: we sketch extensions of LEMMA induced by Jolie, and vice versa.

The common footing we establish brings us one step closer to an ecosystem that coherently combines MDE and programming abstractions to offer a tower of abstractions~\cite{M09} that supports a step-by-step refinement process going from the abstract specification of a microservice architecture (MSA) to its implementation.

\section{A Structured Comparison of Jolie and LEMMA}
\label{sec:comparison}
 
\def\circnum#1{\raisebox{.5pt}{\textcircled{\raisebox{-.9pt}{#1}}}}

\begin{figure}
  \includegraphics[clip, width=\textwidth]{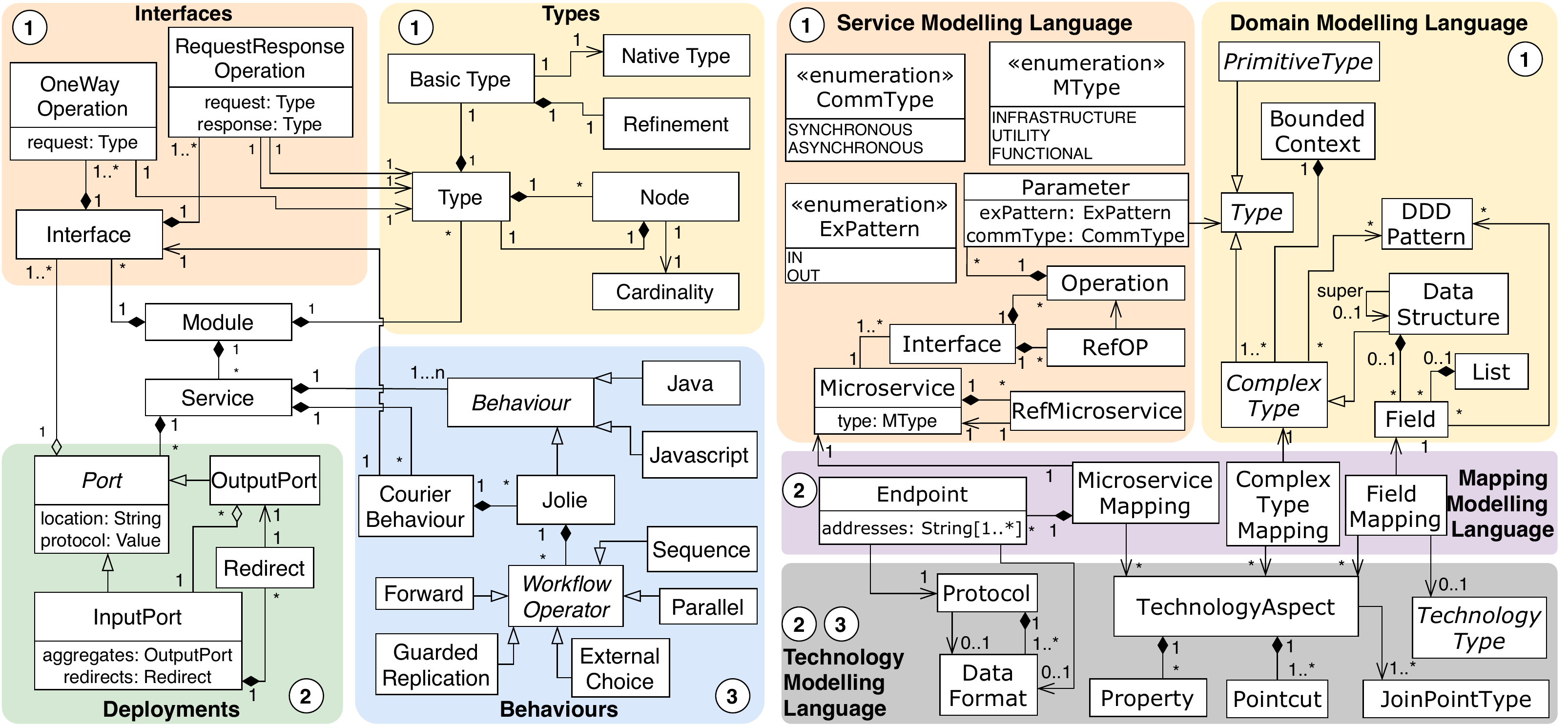}
  \begin{subfigure}{.5\textwidth}
    \smallskip
    \caption{\label{fig:jolie_model}}
  \end{subfigure}  
  \begin{subfigure}{.5\textwidth}
    \smallskip
    \caption{\label{fig:lemma_model}}
  \end{subfigure}
  \vspace{-1.5em}
  \caption{Core meta-models of Jolie (a) and LEMMA (b).}
  \vspace{-1em}
\end{figure}

The conceptual metamodels of Jolie (new in this article) and LEMMA (a
simplification of the metamodel in~\cite{Rademacher2020}) are respectively
displayed in \cref{fig:jolie_model} and \cref{fig:lemma_model}, in UML format.
As a basis for comparison, we classify their elements in three categories:
\hyperref[sec:comparison_api]{Application Programming Interfaces (APIs)
\circnum{1}} and  \hyperref[sec:comparison_access_points]{Access Points
\circnum{2}}, which, combined, define the public contract of a microservice, and
the private internal \hyperref[sec:comparison_behaviours]{behaviour \circnum{3}}
that a microservice enacts. We proceed with explaining the metamodels and our
comparison by following these categories.

\subsection{Application Programming Interfaces (APIs)}
\label{sec:comparison_api}
APIs---originally introduced to provide hardware independence to
programs~\cite{CG68}---specify \emph{what} functionalities a microservice offers
to clients~\cite{Dragoni2017}.
Besides loosing coupling, APIs contribute to technology agnosticism, especially
when minimising the assumptions made on the
technologies used to implement behaviours.

Jolie conceptualises APIs into Interfaces. 
An Interface is a collection of Operations, each having its own name and being either: a OneWay operation, where the sender delivers its message to the service but does not wait for it to be processed by the service's behaviour; or a RequestResponse operation, where the sender delivers its message and waits for
the receiving service's behaviour to reply with a response. 
Operations include types for the data structures that can be exchanged through them. 
A Jolie Type is a tree-shaped data type made of two components: 
\begin{enumerate*}[label={(\roman*)}] 
\item a Basic Type that describes the type of the root of the tree and 
\item a set of Nodes that define the fields of the data structure. Basic Types include a Native Type (primitives like boolean, integer, char, string) and a Refinement that specifies further restrictions on the native type~\cite{FP91}.
\end{enumerate*}
Nodes are arrays with specified ranges of lengths (Cardinality).
Jolie data types, and thus interfaces, are technology agnostic: they model Data Transfer Objects that build on native types generally available in most architectures~\cite{Daigneau2012}.

LEMMA captures APIs as characterising components of a given Microservice though
its Service Modelling Language~\cite{Rademacher2020}. 
Conceptually, a Microservice is a composition of Interfaces, each clustering one or more Operations.
LEMMA distinguishes three types of microservices. Functional and utility ones realise domain-specific business logic and reusable generic functionality, respectively. Infrastructure microservices provide technical capabilities, e.g., for service discovery~\cite{Balalaie2016}.
In LEMMA, a microservice operation is a collection of Parameters, each defined
by an exchange pattern (either incoming or outgoing), a communication type
(synchronous or asynchronous), and a Type, expressed in the Domain Modelling
language.
Types can specify some Domain-Driven Design (DDD) semantics in the form of DDD patterns,
e.g., the Entity pattern~\cite{Evans2004} which defines the identifying traits
of the Type's inhabitants, e.g., a Person with a name and birthdate but uniquely
identified by its social security number.

From the above descriptions---also remarked with the colours of the partitions
in \cref{fig:jolie_model} and \cref{fig:lemma_model}, tagged with
\circnum{1}---APIs are captured similarly in Jolie and LEMMA: they both
attribute a paradigm to each operation, either request-response/synchronous or
notification/asynchronous, although Jolie at the level of operations and LEMMA
at the level of parameters. Types in the two models differ, but, besides LEMMA's
DDD semantics, the differences are mostly technical.
We exploit the vicinity of views on APIs between Jolie and LEMMA to propose in
\Cref{sec:cross_fertilisation} an extension of Jolie that captures DDD patterns
of LEMMA's Types.
At the conceptual level, Jolie and LEMMA interpret API design from different
perspectives. Jolie defines APIs as reusable artefacts, separately from services
(a service can then refer to API definitions). In LEMMA, APIs are part of a
service definition. This difference makes for an interesting point for building
a reference metamodel for microservices, as discussed in
\Cref{sec:cross_fertilisation}.

\subsection{Access Points}
\label{sec:comparison_access_points}
When a microservice implements an API, it must make a technological commitment
on \emph{where} and \emph{how} its clients can interact with the API. Access
points fulfil this need, complementing the public APIs of a microservice with
the specification and configuration of the technologies used to 
\begin{enumerate*}[label={(\roman*)}]
\item format data (how data are structured/marshalled for transmission, e.g., JSON); and
\item transmit data (where microservices can contact each other and how data are transported among them, e.g., an IP address).
\end{enumerate*}
Access points are the main elements that increase coupling between microservices, as providers expect clients to include in their technology stacks the technologies used at providers' access points.

Jolie integrates the Port concept (cf. \cref{fig:jolie_model}) to support access point definition and configuration. A Jolie Port determines the \emph{location} of an access point in the form of a URI~\cite{IETF_URI_2005} and associates it with a \emph{protocol}. Furthermore, a Port clusters one or more Jolie Interfaces, which define the operations available at that access point (and also complete the public contract of the given microservice).

Jolie distinguishes between InputPorts and OutputPorts (cf.
\cref{fig:jolie_model}). InputPorts expose a public contract
to clients
while OutputPorts
define access points used in behaviours (cf. \Cref{sec:comparison_behaviours})
to invoke other microservices. 

LEMMA provides the Endpoint concept (cf. \cref{fig:lemma_model}) to model
locations and technologies of access points, as part of a microservice API. To
cope with \emph{technology heterogeneity} in MSA~\cite{Newman2015}, LEMMA treats
technology information as a dedicated concern in microservice modelling. Indeed,
it provides two modelling languages to \begin{enumerate*}[label={(\roman*)}]
\item organise technology information in
dedicated \emph{technology models}; and 
\item assign this information to service
models within dedicated \emph{mapping models}. 
\end{enumerate*}
In the context of access points, technology models cluster Protocols and DataFormats (cf. \cref{fig:lemma_model}) and make them available to mapping models for determining the technical endpoint
characteristics.

Both Jolie and LEMMA support the specification of inbound access points:
Jolie InputPorts and LEMMA Endpoints include the definition of the
technological choices that define the location and the data formats of access
points.
However, Jolie and LEMMA differ in their description of outbound access points:
\begin{enumerate*}[label={(\roman*)}]
\item Jolie provides OutputPorts to specify, in behaviours, the
interaction with the access points of other microservices.
\item LEMMA provides the RefMicroservice concept to specify dependencies among microservices---LEMMA leaves to model processors how to interpret RefMicroservices, e.g., defining deployment precedence.

\end{enumerate*}

\subsection{Behaviours}
\label{sec:comparison_behaviours}

Behaviours specify the internal business logic of a microservice, including when the microservice can accept requests from clients and when it invokes other microservices.
Jolie allows developers to use Java, JavaScript or Jolie Behaviours to express the behaviour of microservices.
Jolie Behaviours are a fragment of the Jolie Language (herein, Jolie Behavioural Language), where microservice behaviours are first-class citizens that, starting from the basic service invocation, can be composed into complex behaviours via high-level \emph{workflow operators} such as Sequence, Parallel, and Guarded Replication. 
The choice of these operators is rooted in process calculi and the study of core languages for service-oriented computing \cite{GuidiLGBZ06,Montesi2014}.
In this sense, the Jolie Behavioural Language can be regarded as a full-fledged specification language for microservices behaviour and, borrowing LEMMA's conceptual organisation, the Jolie Interpreter as its default technology.

LEMMA does not support (yet) complete specifications of microservice behaviours. However, one can use LEMMA's malleable technology modelling language in this direction, defining a suite of technology aspects for declaring general behaviours (e.g.,~that a microservice is guarded by a circuit breaker) and programming new code generators to produce microservice skeletons.

\section{Cross-Fertilisation and Conclusion}
\label{sec:cross_fertilisation}

The conceptual similarities between Jolie and LEMMA regarding APIs, Access Points and Behaviours identified in this work open the door to cross-fertilisation.

\paragraph{Behaviours in LEMMA} As discussed in \cref{sec:comparison_behaviours}, LEMMA does not support complete and general specifications of microservice behaviours. We propose to extend LEMMA with hosting of languages for programming behaviours like the Jolie Behavioural Language. In general, one can envision a suite of such guest languages that users can select from or extend. The snippet below illustrates a typical instance of this scenario where a programmer extends a microservice specification with a behaviour for \lstinline|operation1|. To this end, the programmer imports a behaviour modelling language and a suitable technology for it, in this case, the Jolie Behavioural Language and the Jolie Interpreter.
\begin{lstlisting}[style=lemma]
import microservices from "example.services" as ExampleServices
import #behaviour_language# from "jolie.behaviour_language" as jolie
import technology from "jolie.technology" as jolie_interpreter

#@behaviour_language#(jolie)
§@technology§(jolie_interpreter)
ExampleServices::org.example.Microservice {
  operation1() #{ /* programmed using the given behavioural language" */ }#
}
\end{lstlisting}

\noindent\looseness=-1
This requires a conceptual and technological infrastructure for language integration in some regards similar to quotation \cite{McCarthy60,CheneyLRW14}: APIs modelled in LEMMA need to be rendered available to the guest language and aspects of behaviour interaction and composition need to be made available to LEMMA. This observation suggests that this integration infrastructure could be founded over the core concepts and behaviour operators for service-oriented programming of process calculi that already constitute the foundation of the Jolie Behavioural Language.

\paragraph{DDD patterns in Jolie}

As mentioned in \Cref{sec:comparison_api}, LEMMA's Types can be augmented with
DDD semantics, i.e., constraints imposed by the domain on data structures.
Equipping Jolie with such a feature can increase its expressiveness in useful
ways, which we discuss briefly below.
To capture DDD patterns in Jolie, we can use comment annotations. For example,
we can express the Entity pattern (cf. \Cref{sec:comparison_api}) via the
annotation \lstinline[style=jolie]{#@entity#} below, which associates the
property \lstinline{identity} of the pattern with two sub-nodes of the Person
type (SSN and country):

\begin{lstlisting}[style=jolie]
/// #@entity# { identity = [ SSN, country ] } 
type Person { SSN: string, country: string( length(3) ), name: string }
\end{lstlisting}

An immediate result is using DDD patterns to improve documentation,
by attaching plain-text explanations of the intended usage of types---in unison
with the additional constraints expressed by refinements (cf.
\lstinline[style=jolie]{length(3)} above). More advanced integrations can
elevate DDD patterns at the level of types, opening the door to runtime and
static utilities. For instance, we can have operations ``governed'' by the
semantics of patterns, e.g., to verify entity equality through a unique
\lstinline{assertEquals} operation that checks equality of the components
defined in the identity annotation of the entity's type. Similarly, patterns can
indicate static constraints on types, e.g., there cannot be two Persons,
identified by SSN and country, whose names differ. Pattern-aware execution
engines can enforce static constraints at runtime, e.g., keeping track of the
(privacy-preserving) ``signature'' of each identified entity and its correlated
immutable values.

\paragraph{Reference Metamodel}
Jolie and LEMMA are in remarkable conceptual proximity despite their distant origins---namely Programming Languages and MDE. 
This close match in their conceptual foundations hints at the existence of a 
reference metamodel for MSAs to be uncovered.
This reference metamodel should identify the main concepts of MSA including
their basic properties and relationships to each other. Furthermore, it should
emerge from the analysis of various existing, yet fragmented bodies of MSA
knowledge ranging from pattern collections, over best practices and reference
solutions for certain challenges in MSA, to more formal approaches like
metamodels for programming and modelling languages.
Recent efforts in the area of software deployment automation~\cite{Wurster2020}
reveal the potential of reference metamodels as they \begin{enumerate*}[label={(\roman*)}]
\item reify and organise knowledge about a specific subject area; 
\item enable the comparison and reasoning about alternative approaches to the same issue; and 
\item allow identification of migration paths and cost estimation for technology choices.
\end{enumerate*}
We believe that a reference metamodel for MSAs would be valuable to organise efforts and unify the great number of ad-hoc solutions for recurring challenges and the heterogeneity of MSAs.

\end{document}